\documentclass[aps,prd,12point,twocolumn,nofootinbib,showpacs,superscriptaddress]{revtex4-1}
\def\theequation{\arabic{section}.\arabic{equation}}
\usepackage{psfrag}
\usepackage{subfigure}
\usepackage{color}
\usepackage{mathrsfs}
\usepackage{graphicx}
\usepackage{amssymb, bm}
\usepackage{amsmath, amsthm}
\usepackage{epstopdf}
\usepackage{hyperref}
\usepackage{enumerate}
\usepackage{longtable}

\newcommand{\be}{\begin{equation}}
\newcommand{\ee}{\end{equation}}

\begin{document}
\def\theequation{\arabic{section}.\arabic{equation}} 
%\preprint{}

\title{Imperfect fluid description of modified gravities}

\author{Valerio Faraoni}
\email[]{vfaraoni@ubishops.ca}
%\thanks{}
\affiliation{Department of Physics and Astronomy and {\em STAR} Research 
Cluster, Bishop's University, 2600 College Street, Sherbrooke, Qu\'ebec, 
Canada J1M~1Z7 }

\author{Jeremy C\^ot\'e}
\email[]{jcote16@ubishops.ca}
%\thanks{}
\affiliation{Department of Physics and Astronomy, Bishop's University, 
2600 College Street, Sherbrooke, Qu\'ebec, 
Canada J1M~1Z7}


%\date{\today}

\begin{abstract}

The Brans-Dicke-like field of scalar-tensor gravity can be described as an 
imperfect fluid in an approach in which the field equations are regarded 
as effective Einstein equations. After completing this approach we 
recover, as a special case, the known effective fluid for a scalar coupled 
nonminimally to the Ricci curvature and we describe the imperfect fluid 
equivalent of $f(R)$ gravity. A symmetry of electrovacuum Brans-Dicke 
gravity is translated into a symmetry of the corresponding effective 
fluid. The discussion is valid for any spacetime geometry.

\end{abstract}

\pacs{}
%\keywords{}

\maketitle

\section{Introduction}
\label{sec:1}
\setcounter{equation}{0}

The 1998 discovery of the present acceleration of the universe with type 
Ia supernovae requires a theoretical explanation. The 
standard cosmological model based on general relativity (GR), {\em 
i.e.}, the 
$\Lambda$ cold dark matter ($\Lambda$CDM) model, requires either an 
incredibly small cosmological constant $\Lambda$ or a dark energy with 
very negative pressure introduced completely {\em ad hoc} 
\cite{AmendolaTsujikawabook}. As an alternative to dark energy, many 
researchers have turned to modifying gravity on cosmological scales. This 
option is far from unrealistic because GR has been tested only in a small 
range of 
regimes \cite{tests}. While many such 
possibilities exist \cite{Padilla}, the class of $f(R)$ theories of 
gravity \cite{CCT} seems by far the most popular. Although this 
class is nothing but old scalar-tensor gravity \cite{BransDicke, ST} in 
disguise, many of its features, related or unrelated to cosmology, 
were only understood in the last decade \cite{reviews}. 

Independent motivation for modifying 
Einstein's theory of gravity comes from high energy physics. Virtually 
every attempt to quantize GR introduces deviations from this theory in the 
form of extra degrees of freedom, higher order derivatives in the field 
equations, higher powers of the curvature in the action, or non-local 
terms. For example, the low-energy limit of the simplest string theory, 
the bosonic string theory, yields Brans-Dicke gravity \cite{BransDicke} 
with Brans-Dicke 
coupling parameter $\omega=-1$ \cite{bosonic}.

In many areas of research, especially in cosmology and in models of 
neutron star and white dwarf interiors, it is common to use 
fluids as the matter source of the Einstein equations. In alternative 
theories of gravity, the more complicated field equations are often recast 
as effective Einstein equations by moving geometric terms other than those 
entering the Einstein tensor $G_{ab} \equiv R_{ab}-g_{ab}R/2$ (where 
$R_{ab}$ and $R\equiv g^{ab}R_{ab}$ are the Ricci tensor and the Ricci 
scalar, respectively) to the right hand side and by regarding them as an effective energy-momentum tensor. This approach has proven 
very  
useful in reducing problems of alternative gravity to known problems of 
GR ({\em e.g.}, Ref.~\cite{Hwang}). But how should one interpret the right 
hand side of 
these field equations? In 
this approach, it makes sense to ask whether this effective 
stress-energy tensor can be formally regarded as a fluid, given that a 
fluid is used so often as the matter source in relativistic physics. 
It is not obvious that this will work, given 
the stress-energy tensor's origin 
and the fact that it is merely an {\em effective} stress-energy tensor. 
However, it turns out to be true \cite{Pimentel89}. In Einstein's theory, an 
effective perfect fluid description can be given for a canonical, 
minimally coupled scalar field $\phi$ \cite{Madsen, Madsen2, myphifluid, 
Semiz, Pilo}, and this fact is well known for special spacetimes, 
such as 
the Friedmann-Lema\^itre-Robertson-Walker (FLRW) spaces used in cosmology. 
Roughly 
speaking, fixing the scalar field potential $V(\phi)$ corresponds to 
prescribing the equation of state of the fluid, but this is not a 
one-to-one correspondence \cite{myAJPpaper, Madsen2, 
BayinCooperstockFaraoni, 
Vikmancomment1, Vikmancomment2}. The effective fluid 
description of more general theories containing a scalar field, such as 
k-essence and special cases of Horndeski gravity, has been worked out in 
detail with respect to cosmological perturbations or to general  
spacetimes \cite{k-essence, Vikmancomment1, Vikmancomment2, 
Christopherson}. In the case of general spacetimes, the 
description of a theory containing a scalar (to which we restrict 
ourselves to) as an effective fluid should not be taken for granted. 
A very interesting non-standard scenario is the one  
of Ref.~\cite{LimSawickiVikman} in which scalar field-fluid elements move 
along timelike geodesics but the pressure is not vanishing, thanks to a 
second scalar field acting as a Lagrange multiplier.  Another possibility 
is 
the higher derivative mimetic dark matter scenario containing an 
effective imperfect fluid which is generated by a scalar 
field \cite{MirzagholiVikman}. Similar to our paper, this effective fluid 
has an energy flow, but contrary to our case, it also has vorticity 
\cite{MirzagholiVikman}.

Here we focus on relatively simple scalar field theories, for which 
the effective fluid description is not yet complete. Among all 
possible 
alternatives to GR, it is natural to first consider scalar-tensor gravity 
\cite{BransDicke, ST}, which only adds a (usually massive) scalar degree 
of freedom, the Brans-Dicke-like scalar $\phi$, to the two massless, 
spin-2  
polarizations contained in the metric tensor $g_{ab} $ and familiar 
from GR. The correspondence between the effective stress-energy tensor of 
$\phi$ and a fluid has been worked out explicitly, first  for 
the case of a nonminimally coupled scalar field \cite{Madsen} and then 
for general scalar-tensor gravity \cite{Pimentel89}. In 
general, the 
corresponding fluid is an imperfect fluid, contrary to the case of a 
minimally coupled scalar, which can always be described as a perfect 
fluid when the scalar field gradient is timelike. However, in special 
spacetimes endowed with symmetries, it may be possible to recover the 
perfect fluid behaviour also for nonminimally coupled scalars 
\cite{BarrosoetalPLB, Culetu}.

Here we extend and complete the correspondence between an (imperfect) 
effective fluid and the Brans-Dicke-like field, we show how a symmetry of 
Brans-Dicke gravity translates into a symmetry of this fluid, 
and we 
apply the discussion to $f(R)$ gravity. We do not 
restrict to special situations such as cosmology or black holes and our 
discussion is valid for general 
spacetime geometries.

Scalar-tensor gravity is described by the Jordan frame 
action (we follow the 
notation of Ref.~\cite{Waldbook} and we use units in which Newton's 
constant $G$ 
and the speed of light $c$ are unity) 
\begin{eqnarray}
S_{ST} &=& \frac{1}{16\pi} \int d^4x \sqrt{-g} \left[ \phi R 
-\frac{\omega(\phi )}{\phi} 
\, \nabla^c\phi \nabla_c\phi -V(\phi) \right] \nonumber\\
&&\nonumber\\
&\, & +S^{(m)} \,, \label{STaction}
\end{eqnarray}
where $\phi>0$ is the Brans-Dicke scalar (approximately 
equivalent to the inverse of the effective gravitational 
coupling strength), the function $\omega(\phi)$ (which was a strictly 
constant parameter in the original Brans-Dicke theory 
\cite{BransDicke}) is the ``Brans-Dicke coupling'', $V(\phi)$ is a 
scalar field potential (absent in the original Brans-Dicke theory), 
whereas $S^{(m)}=\int d^4x \sqrt{-g} \, {\cal L}^{(m)} $ describes the 
matter 
sector. Since our 
task here  regards only the gravitational sector, we will not need to 
specify this ordinary matter.

The (Jordan frame) field equations obtained by varying the 
action~(\ref{STaction}) with 
respect to the inverse metric $g^{ab}$ and to the scalar $\phi$ are 
\cite{BransDicke, ST} 
\begin{eqnarray}
R_{ab} - \frac{1}{2}\, g_{ab} R &=& \frac{8\pi}{\phi} \,  T_{ab}^{(m)} 
\nonumber\\
&&\nonumber\\
&\, & + \frac{\omega}{\phi^2} \left( \nabla_a \phi 
\nabla_b \phi -\frac{1}{2} \, g_{ab} 
\nabla_c \phi \nabla^c \phi \right) \nonumber\\
&&\nonumber\\
&\, &  +\frac{1}{\phi} \left( \nabla_a \nabla_b \phi 
- g_{ab} \Box \phi \right) 
-\frac{V}{2\phi}\, 
g_{ab} \,, \nonumber\\
&& \label{BDfe1} \\
\Box \phi = \frac{1}{2\omega+3} & & 
\left( 
\frac{8\pi T^{(m)} }{\phi}   + \phi \, \frac{d V}{d\phi} 
-2V -\frac{d\omega}{d\phi} \nabla^c \phi \nabla_c \phi \right) \,, 
\nonumber\\
&& \label{BDfe2}
\end{eqnarray}
where $ T^{(m)} \equiv g^{ab}T_{ab}^{(m)} $ is the trace of the matter 
stress-energy tensor 
$T_{ab}^{(m)} $. The matter energy-momentum tensor and the effective 
stress-energy tensor of the scalar $\phi$ are covariantly conserved 
separately. Let us proceed to examine the effective 
fluid description of these field equations and their implications.

\section{Kinematics of the scalar field fluid}
\label{sec:2}
\setcounter{equation}{0}

In this section we identify the kinematic quantities which describe the 
effective fluid 
associated with the  Brans-Dicke-like scalar field. The $\phi$-fluid 
 correspondence is possible when the gradient $\nabla^a \phi$ is timelike;  
then one can introduce  the fluid four-velocity  
\be
u^a  = \frac{\nabla^a  \phi}{\sqrt{ -\nabla^e \phi \nabla_e \phi }}\,, 
\label{4-velocity}
\ee
which is clearly normalized, $u^c u_c=-1$. 
This timelike vector field determines the $3+1$ splitting of spacetime 
into the  3-dimensional space ``seen'' by the comoving observers of the 
fluid and their time direction $u^a$. This 
3-space is endowed with the Riemannian metric 
\be
h_{ab} \equiv g_{ab} + u_a u_b \,,
\ee 
while 
${h_a}^b$ is the usual projection operator on this 3-space and satisfies  
\begin{eqnarray}
h_{ab} u^a &=& h_{ab}u^b=0 \,,\\
&&\nonumber\\
{h^a}_b \, {h^b}_c &=& {h^a}_c \,, \;\;\;\;\;\;  {h^a}_a=3 \,.
\end{eqnarray}
The fluid four-acceleration  
\be
\dot{u}^a \equiv u^b \nabla_b 
u^a 
\ee
is orthogonal to the four-velocity, $\dot{u}^c u_c=0$. 

The (double) projection of the velocity gradient onto the 3-space 
orthogonal to $u^c$ is the purely spatial tensor 
\be
V_{ab} \equiv  {h_a}^c \, {h_b}^d \, \nabla_d u_c  \,, \label{Vab}
\ee 
which is decomposed into its symmetric and 
antisymmetric 
parts, while the symmetric part is further decomposed into its trace-free 
and 
pure trace parts:
\be
V_{ab}=  \theta_{ab} +\omega_{ab} =\sigma_{ab} +\frac{\theta}{3} \, 
h_{ab}+ \omega_{ab} \,,
\ee
where the expansion tensor $\theta_{ab}=V_{(ab)}$ is the symmetric part of 
$V_{ab}$, $\theta\equiv {\theta^c}_c =\nabla^c u_c $ is its trace, the 
vorticity tensor 
$\omega_{ab}=V_{[ab]}$ is its antisymmetric 
part, and the shear tensor
\be 
\sigma_{ab} \equiv \theta_{ab}-\frac{\theta}{3}\, h_{ab}
\ee
is the trace-free part of $\theta_{ab}$. Like $h_{ab}$ and 
$V_{ab}$, expansion, vorticity, and shear 
are purely spatial tensors,
\be
\theta_{ab}u^a = \theta_{ab}u^b = \omega_{ab} \, u^a = \omega_{ab} \, u^b 
=  \sigma_{ab}u^a = \sigma_{ab} u^b = 0 \,,
\ee
and ${\sigma^a}_a={\omega^a}_a=0$ by definition. The shear scalar $\sigma$ 
and the vorticity 
scalar $\omega$ (not to be confused with the Brans-Dicke coupling) are 
defined by 
\begin{eqnarray}
\sigma^2 & \equiv &\frac{1}{2} \, \sigma_{ab}\sigma^{ab} \,,\\
&&\nonumber\\
\omega^2 & \equiv &\frac{1}{2} \, \omega_{ab}\omega^{ab} \,,
\end{eqnarray}
and they are both non-negative. 
In general, we have \cite{Ellis71}  
\be
\nabla_b u_a =   
\sigma_{ab}+\frac{\theta}{3} \, h_{ab} +\omega_{ab} -  \dot{u}_a 
u_b  =V_{ab} -\dot{u}_a u_b \,. \label{ecce}
\ee
The projection of this equation onto the time direction produces 
$\dot{u}_a$, while the projection onto the 3-space orthogonal to 
$u^a$ gives $V_{ab}$.

Let us specialize these general definitions \cite{Ellis71, Waldbook} to 
our particular case. Contrary to the effective stress-energy tensor, the 
corresponding kinematic quantities for the effective fluid 
were not given in Ref.~\cite{Pimentel89}, which only discussed the equivalence of a 
Brans-Dicke field with a fluid.

The definition~(\ref{4-velocity}) of $u^a$ gives  
\be
h_{ab}= g_{ab}-\frac{ \nabla_a\phi \nabla_b \phi}{ 
\nabla^e\phi \nabla_e\phi} 
\ee
and the velocity gradient 
\be
\nabla_b u_a = \frac{1}{ \sqrt{ -\nabla^e\phi \nabla_e \phi}} \left( 
\nabla_a \nabla_b \phi -\frac{  \nabla_a \phi \nabla^c \phi \nabla_b 
\nabla_c \phi}{\nabla^e\phi \nabla_e \phi} \right) \,.
\ee
The acceleration, its norm  
$\dot{u}^a \dot{u}_a$, and its divergence $\nabla_a 
\dot{u}^a$ are 
\begin{widetext}
\begin{eqnarray}
\dot{u}_a &=& \left( -\nabla^e \phi \nabla_e \phi \right)^{-2} 
\nabla^b \phi 
\Big[ (-\nabla^e \phi  \nabla_e \phi)  \nabla_a \nabla_b 
\phi + \nabla^c  \phi \nabla_b \nabla_c \phi \nabla_a \phi \Big] \,, 
\label{acceleration}\\
&&\nonumber\\
\dot{u}^a  \dot{u}_a &=&  (-\nabla^e \phi \nabla_e \phi)^{-3} \left[ 
-\nabla^e \phi  \nabla_e \phi \nabla_b \phi \nabla^d \phi \nabla^b 
\nabla^a \phi  \nabla_d \nabla_a \phi + \left( \nabla^a 
\phi \nabla^b \phi 
\nabla_a \nabla_b \phi \right)^2 \right] \,,\\
&&\nonumber\\
\nabla_a \dot{u}^a & = & (-\nabla^e \phi \nabla_e \phi)^{-2} \left[ 
-\nabla^e \phi \nabla_e \phi  \nabla^b \phi \square \left( \nabla_b \phi 
\right) 
+ \nabla^c \phi \nabla^a  \phi \nabla^b \phi 
\nabla_b \nabla_a 
\nabla_c \phi \right] \nonumber\\
&&\nonumber\\
&\, & + (-\nabla^e  \phi \nabla_e \phi)^{-3} 
\left[ 
(\nabla^e \phi \nabla_e \phi)^2  \nabla^a \nabla^b \phi \nabla_a \nabla_b 
\phi -\nabla^e \phi \nabla_e \phi   \nabla^b \phi \nabla^c \phi 
\nabla_b \nabla_c \phi \square \phi \right.\nonumber\\
&&\nonumber\\
&\, & \left. - 4 \left( \nabla^e \phi 
\nabla_e \phi \right)
\nabla^c \phi \nabla_b \phi \nabla_a  \nabla_c \phi \nabla^b \nabla^a \phi 
+ 4 \left( \nabla^a \phi \nabla^b \phi  \nabla_b \nabla_a \phi \right)^2 
\right] \,.
\end{eqnarray}   \end{widetext}
Using Eqs.~(\ref{acceleration}) and~(\ref{4-velocity}), it is 
straightforward to check explicitly 
that $\dot{u}_c u^c=0$. 
The timelike worldlines  of the fluid elements, with four-tangents $u^a$, 
are  geodesics if and only  if  $\dot{u}_a = 0 $, or 
\begin{equation}
\nabla^e \phi \nabla_{[e} \phi \nabla_{a]}  \nabla_b \phi \nabla^b \phi = 
0  
\end{equation}
and the tensor $V_{ab}$ defined by Eq.~(\ref{Vab})  reduces to
\begin{eqnarray}
V_{ab} &=& \frac{ \nabla_a 
\nabla_b \phi }{ \left( -\nabla^e  \phi \nabla_e \phi \right)^{1/2} }  
\nonumber\\
&&\nonumber\\
&\, & +\frac{ \left( \nabla_a 
\phi \nabla_b  \nabla_c \phi + \nabla_b \phi  \nabla_a \nabla_c \phi 
\right) \nabla^c 
\phi }{ \left( -\nabla^e \phi \nabla_e  \phi \right)^{3/2} } \nonumber\\
&&\nonumber\\
&\, &  + \frac{ \nabla_d \nabla_c \phi 
\nabla^c \phi \nabla^d \phi }{\left( -\nabla^e 
\phi \nabla_e  \phi \right)^{5/2} } \, \nabla_a \phi \nabla_b \phi \,. 
\end{eqnarray}  
The vorticity 
tensor $ \omega_{ab} \equiv  V_{[ab]} $ vanishes identically, 
together  
with the  vorticity scalar  because the fluid four-velocity $u^c$ 
originates 
from a 
gradient (this fact is, of course, consistent with the general 
statement that $\omega=0$ if and only if $\omega_{ab}=0$ \cite{Ellis71}). 
Then we have, for the $\phi$-fluid,  
\be
V_{ab} = \theta_{ab} \,, \;\;\;\;\;
\nabla_b u_a =   
\theta_{ab}   -  \dot{u}_a  u_b \,, 
\ee
the vector field $u^a$ is hypersurface-orthogonal, and the line element 
can be diagonalized in an appropriate coordinate system. There exists a 
family of 3-dimensional hypersurfaces 
$\Sigma$ with Riemannian metric $h_{ab}$ which are orthogonal to the 
four-velocity field $u^a$ and coincide with the 3-spaces seen by 
observers comoving with the fluid, who have four-velocity $u^a$ 
\cite{Ellis71, Waldbook}. 

Since $u^a $ and $\dot{u}^a$ are orthogonal, it is clear from 
Eq.~(\ref{ecce}) that the expansion 
scalar  reduces to the divergence  
\begin{eqnarray} 
\theta = \nabla_a u^a &=& 
\frac{ \square  \phi}{ \left (-\nabla^e \phi 
\nabla_e \phi \right)^{1/2} } \nonumber\\
&&\nonumber\\
&\, &  + \frac{ \nabla_a 
\nabla_b \phi \nabla^a \phi \nabla^b \phi }{ \left( -\nabla^e \phi 
\nabla_e 
\phi \right)^{3/2} } \,. \label{thetaScalar}
\end{eqnarray}
The shear tensor is 
\begin{widetext}
\begin{eqnarray}
 \sigma_{ab} 
&=&  \left( -\nabla^e \phi \nabla_e \phi \right)^{-3/2} \left[ 
-\left( \nabla^e 
\phi \nabla_e 
\phi \right) \nabla_a \nabla_b  \phi 
- \frac{1}{3} \left(  \nabla_a \phi \nabla_b \phi   - g_{ab} \, \nabla^c 
\phi \nabla_c \phi   \right) \square \phi  \right.\nonumber\\
     &&\nonumber\\ 
     &\, & \left. - \frac{1}{3} \left( g_{ab} + \frac{ 2 \nabla_a \phi 
\nabla_b \phi }{   \nabla^e \phi \nabla_e \phi } 
 \right) \nabla_c \nabla_d 
\phi \nabla^d \phi \nabla^c \phi + \left( \nabla_a \phi \nabla_c 
\nabla_b 
\phi + \nabla_b \phi \nabla_c \nabla_a \phi \right) \nabla^c \phi \right]  \,, 
\end{eqnarray}  
while the shear scalar reads
\begin{eqnarray}
\sigma &  = & \left( \frac{1}{2} \, \sigma^{ab} \sigma_{ab} \right)^{1/2} 
 =  ( -\nabla^e \phi \nabla_e \phi)^{-3/2} \left\{ \frac{1}{2} 
\left( \nabla^e  \phi \nabla_e \phi \right)^2 \left[ \nabla^a \nabla^b 
\phi \nabla_a 
\nabla_b \phi  - \frac{1}{3} \left( \square \phi \right)^2 \right]   
\right. \nonumber\\
 &&\nonumber\\
&\, & \left.  +  \frac{1}{3} \left( \nabla_a \nabla_b \phi \nabla^a \phi 
\nabla^b \phi  \right)^2  - \left( \nabla^e \phi \nabla_e 
\phi \right) \left( \nabla_a 
\nabla_b \phi  \nabla^b \nabla_c \phi - \frac{1}{3} \, \square \phi 
\nabla_a 
\nabla_c \phi \right) \nabla^a  \phi \nabla^c \phi \right\}^{1/2} 
\,. \label{sigmaScalar}
\end{eqnarray} 
\end{widetext}
Since $\sigma^2 \ge 0$, Eq.~(\ref{sigmaScalar}) yields the inequality
\begin{eqnarray}
& \, & -\nabla^e \phi \nabla_e \phi  \left[ \nabla^a 
\nabla^b 
\phi \nabla_a \nabla_b \phi - \frac{1}{3} \left(  \square \phi \right)^2 
\right] \nonumber\\
&&\nonumber\\
&+ & 2 \left( \nabla_a \nabla_b \phi \nabla^b  \nabla_c 
\phi - 
\frac{1}{3} \, \square \phi \, \nabla_a \nabla_c \phi \right)  \nabla^a \phi 
\nabla^c \phi  \nonumber\\
&&\nonumber\\ 
&\, & -  \frac{2 \left( \nabla_a \nabla_b \phi \nabla^a \phi 
\nabla^b \phi  \right)^2 }{3 \nabla^e \phi \nabla_e \phi} \ge 0 \,.
\end{eqnarray}
Remember that $\sigma^2=0$ if and only if $\sigma_{ab}=0$ \cite{Ellis71}.  
 In many applications involving the focusing or defocusing of the fluid 
worldlines it is 
 essential to know the sign of the expansion, which is given by 
Eq.~(\ref{thetaScalar}):
\begin{equation}
     \theta \ge 0 \iff 
 \nabla^a \phi \nabla^b \phi \nabla_a \nabla_b \phi  -\nabla^e \phi 
\nabla_e \phi  \square \phi  \ge 0 \,.
\end{equation}

\section{Scalar field effective stress-energy tensor}
\label{sec:3}
\setcounter{equation}{0}

In scalar-tensor gravity, the effective stress-energy tensor of the 
Brans-Dicke-like field is given by
\begin{eqnarray}
8\pi T_{ab}^{(\phi)} &=& \frac{\omega}{\phi^2} \left( \nabla_a \phi 
\nabla_b \phi - 
 \frac{1}{2} \, g_{ab} \nabla^c \phi \nabla_c \phi  \right) \nonumber\\
&&\nonumber\\
&\, & + 
 \frac{1}{\phi} \left( \nabla_a \nabla_b \phi -g_{ab} \square \phi \right) 
- \frac{V}{2 \phi} \, g_{ab} \,. \label{BDemt}
\end{eqnarray}
This effective stress-energy tensor and the matter stress-energy tensor 
are 
covariantly conserved separately,
\be
\nabla^b T_{ab}^{(m)}=0 \,, \;\;\;\;\;
\nabla^b T_{ab}^{(\phi)}=0 \,.
\ee
It was shown in Ref.~\cite{Pimentel89} that the  effective energy-momentum 
tensor $T_{ab}^{(\phi)}$  can be  written  
as the stress-energy tensor of an imperfect fluid. The latter
admits the decomposition 
\be 
T_{ab} = \rho u_a u_b + q_a u_b + q_b u_a + \Pi_{ab} 
\,,\label{imperfectTab}
\ee 
where 
\begin{eqnarray}
\rho &=& T_{ab} u^a u^b \,,  \label{rhophi}\\
&&    \nonumber\\
q_a & =&  -T_{cd} \, u^c {h_a}^d \,, \label{qphi}\\
&&  \nonumber\\
 \Pi_{ab} &\equiv & Ph_{ab} + \pi_{ab} = T_{cd} \, {h_a}^c \, {h_b}^d \,, 
\label{Piphi}\\
&&  \nonumber\\
    P &=& \frac{1}{3}\, g^{ab}\Pi_{ab} =\frac{1}{3} \, h^{ab} T_{ab} \,, 
\label{Pphi}\\
&&  \nonumber\\
    \pi_{ab} &=& \Pi_{ab} - Ph_{ab} \,, \label{piphi}
\end{eqnarray}
are,  respectively, the effective energy density, heat flux density, 
stress tensor,\footnote{The right hand side in the last equality 
of~(\ref{Piphi}) points to an incorrect sign in Eq.~(9) 
of Ref.~\cite{Pimentel89}, but this incorrect sign was not used in the 
rest of this reference.}  
isotropic  pressure, and anisotropic stresses (the trace-free part 
$\pi_{ab}$ of 
the stress tensor $\Pi_{ab}$) in the comoving frame. In 
this frame, by definition, the fluid elements are at rest and the 
heat flux density, which is the only energy flow, is purely 
spatial,  
\be 
q_c u^c = 0 \,,
\ee 
while 
\be 
\Pi_{ab} u^b=\pi_{ab} u^b = \Pi_{ab} u^a=\pi_{ab} u^a = 0 \, , 
\,\,\,\,\,\,\,  {\pi^a}_a = 0 \,. 
\ee 
The covariant conservation of $T_{ab}^{(\phi)}$ can be projected along the 
time direction $u^a$ and on the 3-space with Riemannian metric $h_{ab}$, 
which yields 
\cite{Ellis71}
\begin{eqnarray}
&& u^a\nabla_a \rho^{(\phi)} +\left( P^{(\phi)} +\rho^{(\phi)} 
\right) \theta +\Pi^{ab} \sigma_{ab} +\nabla^a q_a +q^a \dot{u}_a \nonumber\\
&& =0 \,,\\
&&\nonumber\\
&& \left( P^{(\phi)} +\rho^{(\phi)} \right) \dot{u}_a +{h_a}^c \left( 
\nabla_c P^{(\phi)} + \nabla^b \Pi_{cb} + u^e\nabla_e q_c \right) 
\nonumber\\
&&\nonumber\\
&& + \left( {\omega_a}^b + {\sigma_a}^b +\frac{4}{3} \, \theta \, {h_a}^b 
\right) 
q_b =0 \,.
\end{eqnarray}
Since we used the fluid four-velocity $u^c$ and the corresponding 3-metric 
$h_{ab}$ to project, the imperfect fluid quantities defined in this way 
are those in the comoving frame. When  calculated explicitly, they read 
\cite{Pimentel89} 
\begin{widetext}
\begin{eqnarray}
8 \pi \rho^{(\phi)} &=&  -\frac{\omega}{2\phi^2} \, \nabla^e \phi \nabla_e 
\phi  +  \frac{V}{2\phi} + \frac{1}{\phi} \left( \square \phi -  
\frac{  \nabla^a \phi \nabla^b \phi \nabla_a 
\nabla_b \phi}{ \nabla^e \phi  \nabla_e \phi  } \right)  
\,,\label{effdensity}\\
&&\nonumber\\
8 \pi q_a^{(\phi)}   &=& \frac{\nabla^c  \phi \nabla^d \phi}{\phi 
  \left(-\nabla^e \phi \nabla_e \phi \right)^{3/2} } \,  
\Big(  \nabla_d \phi \nabla_c \nabla_a \phi 
- \nabla_a \phi \nabla_c \nabla_d \phi \Big)  \label{eq:q}\\
&&\nonumber\\
&=& -\frac{\nabla^c \phi \nabla_a \nabla_c\phi}{\phi \left( -\nabla^e \phi 
\nabla_e\phi \right)^{1/2}} - \frac{
\nabla^c\phi \nabla^d\phi \nabla_c\nabla_d \phi}{\phi \left( -\nabla^e\phi 
\nabla_e\phi \right)^{3/2} } \, \nabla_a \phi \,,\\
&&\nonumber\\
8 \pi \Pi_{ab}^{(\phi)}  &=&  (-\nabla^e  \phi \nabla_e \phi)^{-1} \left[ 
\left( 
- \frac{\omega}{2\phi^2}\,  \nabla^e \phi \nabla_e \phi - \frac{\square 
\phi}{\phi} - \frac{V}{2 \phi} \right) \Big( \nabla_a \phi \nabla_b \phi -
g_{ab}  \nabla^e \phi  \nabla_e \phi  \Big) \right.  
\nonumber\\
&&\nonumber\\
&\, & \left.  - \frac{\nabla^d 
\phi}{\phi} \left( \nabla_d  \phi \nabla_a \nabla_b \phi - \nabla_b \phi 
\nabla_a \nabla_d \phi - \nabla_a  \phi \nabla_d \nabla_b \phi + 
\frac{  \nabla_a \phi \nabla_b \phi \nabla^c \phi  \nabla_c \nabla_d 
\phi}{ \nabla^e \phi \nabla_e \phi } \right) \right] \\
&&\nonumber\\
&=& \left( -\frac{\omega}{2\phi^2} \, \nabla^c \phi \nabla_c \phi 
-\frac{\Box\phi}{\phi} -\frac{V}{2\phi} \right) h_{ab} +\frac{1}{\phi} \, 
{h_a}^c {h_b}^d \nabla_c \nabla_d \phi \,,  \label{eq:effPi2}\\
&&\nonumber\\
8 \pi P^{(\phi)}  & = &  - \frac{\omega}{2\phi^2} \, \nabla^e \phi 
\nabla_e \phi - 
\frac{V}{2\phi} - \frac{1}{3\phi}  \left( 2\square \phi + 
\frac{\nabla^a \phi \nabla^b \phi \nabla_b \nabla_a \phi }{\nabla^e \phi 
\nabla_e  \phi }  \right) \,, \label{effpressure}\\
&&\nonumber\\
8 \pi \pi_{ab}^{(\phi)}   &=& \frac{1}{\phi \nabla^e \phi \nabla_e 
\phi } 
\left[ \frac{1}{3} \left( \nabla_a  \phi \nabla_b \phi - g_{ab} \nabla^c 
\phi \nabla_c \phi \right) \left(  \square \phi  - 
\frac{ \nabla^c \phi  \nabla^d \phi \nabla_d \nabla_c \phi }{ \nabla^e \phi 
\nabla_e \phi }   
\right) \right. \nonumber\\
&&\nonumber\\
&\, & \left. + \nabla^d \phi \left(  \nabla_d \phi \nabla_a \nabla_b 
\phi - 
\nabla_b \phi \nabla_a \nabla_d  \phi - \nabla_a \phi \nabla_d \nabla_b 
\phi +  
\frac{ \nabla_a \phi \nabla_b \phi  \nabla^c \phi \nabla_c 
\nabla_d \phi }{ \nabla^e \phi \nabla_e \phi } \right) \right] \,, 
\label{piab-phi}\\
&&\nonumber\\
8 \pi T^{(\phi)}  & \equiv &  8 \pi g^{ab}T_{ab}^{(\phi)}  
 = - \frac{\omega}{ \phi^2} \, \nabla^c \phi 
\nabla_c  \phi - \frac{3\square \phi}{\phi} - \frac{2V}{\phi} 
\,.\label{eq:efftrace}
\end{eqnarray} 
\end{widetext}
Apart from different notations, 
Eqs.~(\ref{effdensity})-(\ref{eq:efftrace}) 
agree with the corresponding expressions of 
Ref.~\cite{Pimentel89}. It is straightforward to check explicitly that 
$q^a, \Pi^{ab}, 
\pi^{ab}$ are purely spatial and that the trace of the effective 
stress-energy tensor of $\phi$ coincides, as it should be, with $ 
-\rho^{(\phi)}   + 
3P^{(\phi)}  $, with $\rho^{(\phi)} $ and $P^{(\phi)}  $ given by 
Eqs.~(\ref{effdensity}) and~(\ref{effpressure}). Furthermore, by  
comparing Eqs.~(\ref{eq:q}) and (\ref{acceleration}), one obtains
\be
q_a^{(\phi)} =  -\frac{ \sqrt{-\nabla^c \phi \nabla_c \phi}}{ 8 \pi \phi} \, 
\dot{u}_a \,,\label{q-a}  
\ee 
making obvious the fact that this vector is purely spatial. What is more, 
it is easy to see that in general the heat flux density $q_a^{(\phi)}$ and 
the anisotropic stresses $\pi_{ab}^{(\phi)}$ do not vanish and that the 
effective fluid is necessarily an imperfect one. The relation~(\ref{q-a}), 
which seems to have gone unnoticed in the literature thus far, has a 
physical consequence in Eckart's first order 
thermodynamics \cite{Eckart40} (which is notoriously plagued by 
non-causality and instability but is still widely used as an 
approximation). In this theory, the heat flux density is related to the 
temperature ${\cal T}$ by the generalized Fourier law \cite{Eckart40} 
\begin{eqnarray}
 q_a &=& -K \left( h_{ab} \nabla^b {\cal T} +{\cal T} \dot{u}_a \right) 
\,, \label{Eckart}\\
&&\nonumber
\end{eqnarray}
where $K$ is the thermal conductivity. The comparison of Eqs.~(\ref{q-a}) 
and (\ref{Eckart}) leads to the result that, in the comoving frame, the 
spatial temperature gradient vanishes and the heat flow is then due purely 
to the inertia of energy described by the acceleration term in 
Eq.~(\ref{Eckart}). Moreover, the product of the thermal conductivity and 
the temperature of the effective fluid is
\be
K{\cal T}= \frac{ \sqrt{-\nabla^c \phi \nabla_c\phi}}{8\pi \phi} \,,
\ee
which is positive definite.

An alternative approach consists of trading temperature with chemical 
potential, assigning zero temperature and entropy but nonzero chemical 
potential to the effective fluid. This approach is pioneered in 
Ref.~\cite{Vikmancomment2} (note the similarity between our 
Eq.~(\ref{q-a}) and Eq.~(3.35) of Ref.~\cite{Vikmancomment1}).

It is sometimes convenient to replace the d'Alembertian  $\square  \phi$ 
in  the  expressions of $\rho^{(\phi)} $ and $P^{(\phi)} $ 
with its value obtained from the 
field equation~(\ref{BDfe2}). For reference, we provide the corresponding 
expressions
\begin{widetext}
\begin{eqnarray}
8 \pi \rho^{(\phi)}   &=& - \frac{\omega}{2\phi^2} \, \nabla^e  \phi 
\nabla_e 
\phi + \frac{V}{2\phi} \left( \frac{2\omega -1}{2  \omega + 3} \right) 
\nonumber\\
&&\nonumber\\
&\, & + 
\frac{1}{\phi} \left[ \frac{1}{2\omega + 3}  \left( \phi \, 
\frac{dV}{d\phi} -  
\nabla^e \phi \nabla_e \phi \, \frac{d\omega}{d\phi}  \right) - 
\frac{  \nabla^a \phi \nabla^b \phi \nabla_a \nabla_b \phi}{ 
\nabla^e \phi \nabla_e \phi} \right] 
\,, \label{rhophi2}\\
&&\nonumber\\
8 \pi P^{(\phi)}  &=&  - \frac{\omega}{2\phi^2} \, \nabla^e \phi  \nabla_e 
\phi -  \frac{V}{6\phi} \, \frac{ \left(6\omega + 1\right)}{\left(2\omega  
+  3\right)} \nonumber\\
&&\nonumber\\
&\, &  -  \frac{1}{3\phi} \left[ \frac{2}{2\omega+3} \Big( \phi \, 
\frac{dV}{d\phi} - 
  \nabla^e \phi \nabla_e \phi \, \frac{d\omega}{d\phi} \Big) + \frac{ 
\nabla^a \phi 
\nabla^b \phi \nabla_b \nabla_a \phi }{\nabla^e 
\phi  \nabla_e \phi }  \right] \,. \label{Pphi2}
\end{eqnarray}  
\end{widetext}
It is now straightforward to check that the imperfect fluid stress-energy 
tensor~(\ref{imperfectTab}) is reproduced by $T_{ab}^{(\phi)}$. In fact, 
adding Eqs.~(\ref{effdensity})-(\ref{piab-phi}) with the appropriate 
coefficients, one obtains
\be
\rho^{(\phi)} u_a u_b +q^{(\phi)}_a u_b + q^{(\phi)}_b u_a + 
\Pi^{(\phi)}_{ab} = T_{ab}^{(\phi)} \,.
\ee
Being built by hand out of geometric or 
gravitational terms, the effective 
fluid stress-energy tensor $T_{ab}^{(\phi)} $, in general, does not satisfy any 
energy condition because of the presence of second derivatives of $\phi$. 
The weak energy condition $T_{ab} t^a t^b \geq 0$ for all timelike vectors 
$t^a$ \cite{Waldbook} becomes, for $T_{ab}^{(\phi)} $ and 
the fluid four-velocity,
\begin{eqnarray}
&& -\frac{\omega}{2\phi} \, \nabla^e \phi \nabla_e  \phi + 
\frac{V}{2} +  \square \phi   -
\frac{ \nabla^a \phi \nabla^b \phi \nabla_a  \nabla_b \phi }{\nabla^e \phi 
\nabla_e \phi } \geq 0 \,.\nonumber\\
&&
\end{eqnarray}
The strong energy condition $    \left( T_{ab} -  T g_{ab}/2 
\right) t^a t^b  \geq 0$ for all timelike vectors $t^a$ 
\cite{Waldbook} reads, when 
applied to $u^a$ and to $T_{ab}^{(\phi)} $,
\begin{eqnarray}
&&    \left( T_{ab}^{(\phi)}  - \frac{1}{2}\, T^{(\phi)}   g_{ab} \right) 
u^a u^b =  \frac{1}{2} 
\left( \rho^{(\phi)}  + 3P^{(\phi)}  \right) \nonumber\\
&&\nonumber\\
&&= - \frac{\omega}{\phi^2} \, \nabla^e  
\phi \nabla_e 
\phi - \frac{V}{2\phi} \nonumber\\
&&\nonumber\\
&\, & + \frac{1}{\phi} \left[ -\frac{1}{2}  
\square \phi - \frac{ \nabla^a \phi \nabla^b  \phi 
\nabla_a \nabla_b \phi}{ \nabla^e \phi \nabla_e \phi }  \right] \ge 0 
\,.  
\end{eqnarray}
However, as noted above, it is not physically meaningful to impose these 
energy conditions on $T_{ab}^{(\phi)}$. Moreover, the energy conditions 
reported involve density and stresses {\em in the comoving frame} because 
we imposed that the observer coincides with the fluid four-velocity ({\em 
i.e.},  $t^a=u^a$). The energy conditions of an imperfect fluid with 
respect to arbitrary timelike observers are discussed in 
\cite{Kolassisetal88, Pimenteletal6}.

\section{A special case: the nonminimally coupled scalar field}
\label{sec:4}
\setcounter{equation}{0}

The correspondence between a Brans-Dicke-like scalar $\phi$ and an 
imperfect fluid was studied in Ref.~\cite{Madsen}, for general spacetimes, 
in the case of a 
scalar field 
coupling nonminimally to the Ricci curvature. Such a nonminimal coupling 
appears when quantizing a canonical, minimally coupled test 
scalar field in a curved space \cite{CCJ1} and also in the context of 
radiation problems \cite{ChernikovTagirov, DeWittBrehme, SonegoFaraoni} 
(see also Refs.~\cite{Odintsov5, CCJ2, CCJ3, CCJ4, Friedlander}). The 
nonminimal 
coupling of the scalar has been studied extensively during  inflation of 
the early universe (\cite{NMCinflation} and references therein).
When the scalar is allowed to 
gravitate, one has, for all practical purposes, a scalar-tensor theory 
\cite{FujiiMaeda, mySTbook}.  We now show that a known representation of 
the theory of a nonminimally 
coupled scalar field as an imperfect fluid \cite{Madsen}  
is contained, as a special case, in our 
formulas. The action 
for  a nonminimally coupled scalar $\phi$ is
\begin{eqnarray}
S_{NMC} &=& \int d^4x \sqrt{-g} \Big[ \left( \frac{1}{8\pi} -  \xi  
\phi^2 \right) \frac{R}{2} - \frac{1}{2}\, \nabla^e \phi 
\nabla_e \phi  \nonumber\\
&&\nonumber\\
&\, &  - V(\phi) \Big] \,,\label{NMCaction}
\end{eqnarray}
where $\xi$ is the dimensionless coupling constant (with $\xi=1/6$ 
corresponding to conformal coupling \cite{CCJ1, Waldbook}), the value of 
which depends on the nature of the scalar and can often be determined 
as a running coupling going to an infrared fixed point under a 
renormalization group 
flow \cite{xi-value, Odintsov5}.  
Our general scalar-tensor action {\em in vacuo} is instead 
\begin{equation}
S_{ST} = \frac{1}{16\pi} \int d^4x \sqrt{-g} \left[ \psi R - 
\frac{\omega(\psi)}{\psi} \,\nabla^c \psi \nabla_c \psi  - V(\psi) 
\right] \,.\label{ourSTaction}
\end{equation}
In order to establish the connection between these two actions, it is 
sufficient to write the particular form of the  
function  $\omega (\psi) $ that corresponds to the 
Brans-Dicke-like representation of the action~(\ref{NMCaction}). 
Identifying the first term in each action gives
\begin{equation}
    \psi = 1 - 8\pi \xi\phi^2  \,.
\end{equation}
Contrary to the Brans-Dicke-like field $\phi$, the nonminimally coupled  
scalar $\phi$ is not restricted to be positive. However, since $\psi>0$, 
for $\xi>0$ the scalar $\phi$ must satisfy $|\phi|< 
\phi_c \equiv 1/\sqrt{ 8\pi \xi}$, while all values of $\phi$ are 
admissible if $\xi<0$. By using
\begin{eqnarray}
\phi & = & \pm \sqrt{ \frac{1-\psi}{ 8\pi \xi}}  \,, \\
&&\nonumber\\
\nabla_e \phi & = & \mp  \frac{ \nabla_e \psi }{\sqrt{32\pi \xi 
\left(1 - \psi\right) }} \,, \\
&&\nonumber\\
\nabla_a  \nabla_b (\phi^2) &=& -\frac{1}{8\pi \xi} \, \nabla_a \nabla_b 
\psi \,, 
\end{eqnarray}
one finds easily
\be
 \omega(\psi)  = \frac{\psi}{4 \xi \left(1 - \psi \right) } \,.
\ee
{\em Vice-versa}, it is 
\begin{eqnarray} 
\nabla_e \psi & = & -16\pi \xi \phi \nabla_e \phi \,, \\  
&&\nonumber\\
\nabla_a \nabla_b \psi &=& -8\pi\xi \,\nabla_a \nabla_b (\phi^2) \,, \\ 
&&\nonumber\\
\omega\left( \psi(\phi) \right) &= & \frac{ 1-8 \pi \xi \phi^2 }{
32\pi\xi^2 \phi^2} \,.
\end{eqnarray}
These transformation properties allow us to recover  the 
theory described by the action~(\ref{NMCaction}), the effective fluid 
representation of which is discussed in Ref.~\cite{Madsen}, as a special 
case of   
the general theory described by the scalar-tensor 
action~(\ref{ourSTaction}). This limit was not derived in 
Ref.~\cite{Pimentel89}. 

The effective stress-energy tensor~(\ref{BDemt}) of the Brans-Dicke-like 
scalar field $\psi$ becomes, in terms of the new field $\phi$, 
\begin{eqnarray}
T_{ab}^{(\psi)} &=& \left( 1-8\pi\xi \phi^2 \right)^{-1} 
\Big\{  \nabla_a \phi \nabla_b \phi -\frac{1}{2} \, g_{ab} \nabla^e \phi 
\nabla_e \phi  \nonumber\\
&&\nonumber\\
& \, &  -\frac{U(\phi)}{2} \, g_{ab} -\xi \left[
\nabla_a\nabla_b(\phi^2) -g_{ab} \square (\phi^2)  \right] \Big\} \,,
\nonumber\\ 
&&\label{eq:psiTab}
\end{eqnarray}
where
\be
U(\phi) = \frac{ V\left[ \psi(\phi) \right]}{8\pi} \,.
\ee
As already shown, the energy-momentum tensor~(\ref{eq:psiTab}) assumes the 
form of an imperfect fluid energy-momentum tensor.  The effective energy 
density is 
\begin{widetext}
\begin{eqnarray}
\rho^{(\psi)} &=&  \left(1-8\pi\xi\phi^2\right)^{-1}  
\Bigg\{ -\frac{1}{2}\, \nabla^e \phi \nabla_e \phi 
+ \frac{U(\phi)}{2} + \xi \left[    
\frac{ \nabla^a \phi \nabla^b \phi \nabla_a \nabla_b (\phi^2)}{ \nabla^e 
\phi \nabla_e \phi} -  \square (\phi^2) \right] \Bigg\} \,,
\end{eqnarray}
while the effective heat flux density is 
\begin{eqnarray}
q_a^{(\psi)} &=& \xi \left( 1-8\pi \xi \phi^2 \right)^{-1}  \frac{ 
\nabla^c \phi \nabla^d \phi}{ \left( -\nabla^e \phi\nabla_e 
\phi\right)^{3/2} }  \Big[ 
\nabla_d \phi \nabla_a \nabla_c ( \phi^2)  - \nabla_a \phi  
\nabla_c \nabla_d (\phi^2)  \Big] \,.
\end{eqnarray}
The effective pressure reads 
\begin{eqnarray}
P^{(\psi)} &=& \left(1 - 8\pi \xi \phi^2\right)^{-1} \Bigg\{ 
-\frac{1}{2} \, \nabla^e \phi \nabla_e \phi - 
\frac{U(\phi)}{2} + \frac{\xi}{3} \left[ 2\square (\phi^2) + 
\frac{ \nabla^a \phi \nabla^b \phi \nabla_a \nabla_b (\phi^2)}{\nabla^e 
\phi \nabla_e \phi}  \right] \Bigg\} \,,
\end{eqnarray}
while the stress tensor and the anisotropic stresses are 
\begin{eqnarray}
\Pi_{ab}^{(\psi)} &=&
\left( 1-8\pi\xi \phi^2 \right)^{-1} \Bigg\{
\left[-\frac{1}{2} \,  \nabla^e \phi \nabla_e \phi -\frac{U(\phi)}{2} 
\right] h_{ab} 
+\xi \left[  h_{ab} \square(\phi^2) - {h_a}^c {h_b}^d \nabla_c \nabla_d 
(\phi^2) \right] \Bigg\}  
\end{eqnarray}
and
\begin{eqnarray}
\pi_{ab}^{(\psi)} &=& -\frac{ \xi \left(1- 
8\pi\xi\phi^2\right)^{-1}}{\nabla^e\phi \nabla_e\phi}   \Bigg\{ 
\frac{1}{3} \left( \nabla_a \phi \nabla_b \phi -g_{ab} \nabla^e \phi 
\nabla_e \phi\right) \left[
\square(\phi^2)- \frac{  \nabla^c \phi \nabla^d \phi \nabla_c \nabla_d 
(\phi^2)}{  \nabla^e \phi \nabla_e \phi}\right] \nonumber\\
&&\nonumber\\
&\, & + \nabla^d \phi \left[ \nabla_d \phi  \nabla_a\nabla_b (\phi^2)  
- \nabla_b \phi \nabla_a\nabla_d (\phi^2) + \frac{
\nabla^a \phi \nabla_b \phi \nabla_c \nabla_d (\phi^2)}{
\nabla^e \phi \nabla_e \phi} \right] \Bigg\} \,,
\end{eqnarray}
\end{widetext}
respectively. Finally, the trace of the stress-energy tensor is
\begin{eqnarray}
T^{(\psi)} &=& \left(1-8\pi\xi\phi^2 \right)^{-1} \left[
- \nabla^e \phi \nabla_e \phi - 2U(\phi)   +3\xi \square (\phi^2)  \right] 
\,. \nonumber\\
&&\label{eq:psiT}
\end{eqnarray}
Equations~(\ref{eq:psiTab})-(\ref{eq:psiT}) reproduce the 
corresponding effective fluid quantities of 
Ref.~\cite{Madsen} after accounting for the different notations.

\subsection{Minimally coupled scalar field}

By setting the coupling constant $\xi$ to zero, the scalar $\phi$ 
decouples from the Ricci curvature and assumes the ordinary 
non-gravitational form considered in GR. In the limit $\xi \rightarrow 
0$, Eqs.~(\ref{eq:psiTab})-(\ref{eq:psiT}) yield the effective fluid 
quantities 
\begin{eqnarray}
T_{ab}^{(0)} &=& \nabla_a \phi \nabla_b \phi -\frac{1}{2} \, g_{ab} 
\nabla^e \phi  \nabla_e \phi  -\frac{U(\phi)}{2} \, g_{ab}  \,,\\
&&\nonumber\\
\rho^{(0)} &=&  -\frac{1}{2}\, \nabla^e \phi \nabla_e \phi + 
\frac{U(\phi)}{2} \,,\\
&&\nonumber\\
P^{(0)} &=& -\frac{1}{2} \, \nabla^e \phi \nabla_e \phi - 
\frac{U(\phi)}{2}  \,,\\
&&\nonumber\\
\Pi_{ab}^{(0)} &=& \left[-\frac{1}{2} \,  \nabla^e \phi \nabla_e \phi 
-\frac{U(\phi)}{2}  \right] h_{ab} \,,\\
&\nonumber\\
T^{(0)} &=& - \nabla^e \phi \nabla_e \phi - 2U(\phi)  \,,
\end{eqnarray}
while the heat flux density $q_a^{(0)}$ and the anisotropic stresses 
$\pi_{ab}^{(0)}$ 
vanish identically, giving the energy-momentum tensor  of the minimally 
coupled scalar field the structure of a perfect fluid, as is well known 
\cite{Madsen, Madsen2, myphifluid, Semiz, Pilo}.  One can also write  
\be
T_{ab}^{(0)}=\nabla_a\phi \nabla_b\phi -{\cal L}^{(0)} g_{ab} \,,
\ee
where 
\be
{\cal L}^{(0)}= \frac{1}{2} \, \nabla^c \phi \nabla_c \phi -U(\phi) 
\ee
is the minimally coupled scalar field Lagrangian density. 
Furthermore, it is
\be
{\cal L}^{(0)}=P^{(0)} \,. \label{L=P}
\ee
In the effective fluid approach to scalar field theories, this 
equation is significant because it is consistent with 
the  fact, well known in relativistic and non-relativistic fluid dynamics,  
that equivalent Lagrangian densities 
for a perfect fluid are ${\cal L}_1=P $ and ${\cal 
L}_2=-\rho$ 
\cite{SeeligerWhitham, Brown, Schutz}. These two 
Lagrangians  become  inequivalent if the fluid couples to another 
component of the matter sector of the theory, as discussed in the recent 
Ref.~\cite{myPRDpaper}. When $\phi$ couples to the Ricci curvature ({\em 
i.e.}, $\xi \neq 0$), instead, the equivalent effective fluid is no 
longer a perfect fluid and Eq.~(\ref{L=P}) no longer holds 
(therefore, it is meaningless to discuss the equivalence of $P$ and 
$-\rho$ as Lagrangians).

For the minimally coupled scalar, we have also
\be
\rho^{(0)} =  {\cal L}^{(0)}+2U(\phi) \,.
\ee
An equation of state for this perfect fluid is specified by giving two 
relations $\rho=\rho \left( {\cal L}^{(0)}, U \right) \,,  
P= P \left( {\cal L}^{(0)}, U \right) $ \cite{Madsen2}.

\section{Fluid symmetry for electrovacuum Brans-Dicke gravity}
\label{sec:5}
\setcounter{equation}{0}

Consider now the vacuum or electrovacuum Brans-Dicke theory with 
$\omega=$~const. The Brans-Dicke action~(\ref{STaction}) with constant 
$\omega$  is invariant in form under the 
one-parameter group of symmetries \cite{BDlimit} 
\begin{eqnarray}
&& g_{ab}  \rightarrow  \tilde{g}_{ab}=\phi^{2\alpha} \, g_{ab} \,, 
\label{newg}\\
&&\nonumber\\ 
&& \phi \rightarrow \tilde{\phi} = \phi^{1-2\alpha} \,, \;\;\;\;\;\; 
\alpha\neq 0, 1/2 \,,\label{newphi}
\end{eqnarray}
provided that the Brans-Dicke parameter $\omega$ and the scalar potential 
$V(\phi)$ are replaced by  
\begin{eqnarray}
\tilde{\omega}( \omega, \alpha)&=& \frac{ \omega 
+6\alpha(1-\alpha)}{(1-2\alpha)^2} \,, \label{newomega}\\ 
&&\nonumber\\
\tilde{V}( \tilde{\phi}) &=& \tilde{\phi}^{\frac{-4\alpha}{1-2\alpha}} 
V\left( \tilde{\phi}^{\frac{1}{1-2\alpha}} \right) \,. \label{newV}
\end{eqnarray}
This one-parameter symmetry group is used to generate new solutions from 
known ones \cite{ValerioDilekShawn} and to study the limit to GR of 
Brans-Dicke gravity \cite{BDlimit}. Assuming that $\nabla^c \phi$ 
is timelike, under the 
transformation~(\ref{newg}) and (\ref{newphi})  the 
fluid four-velocity is mapped to
\begin{eqnarray}
u_c & \rightarrow & \tilde{u}_c \equiv \frac{ \tilde{\nabla}_c  
\tilde{\phi} }{ 
\sqrt{ - \tilde{g}^{cd} \tilde{\nabla}_c \tilde{\phi} \tilde{\nabla}_d 
\tilde{\phi} }}       
=\phi^{\alpha} u_c \,, \label{u-property}\\
&&\nonumber\\
u^c & \rightarrow & \tilde{u}^c=\phi^{-\alpha} u^c \,. \label{u2}
\end{eqnarray}
Since $\phi>0$, the transformation property~(\ref{u-property}) preserves 
the timelike 
character of the four-velocity and, since $u^c$ and 
$\tilde{u}^c$ are normalized with respect to different 
metrics, it also preserves the normalization of the four-velocity, 
\be
\tilde{g}^{ab} \tilde{u}_a  \tilde{u}_b  
= g^{ab} u_a u_b = -1 \,.
\ee
The fluid quantities $\tilde{\rho}^{( \tilde{\phi})} , 
\tilde{P}^{( \tilde{\phi})}, \tilde{q}_a^{( \tilde{\phi})},
\tilde{\Pi}_{ab}^{( \tilde{\phi})}$, and $ 
\tilde{\pi}_{ab}^{( \tilde{\phi})} $ are given in terms of $\tilde{\phi}$ 
and its derivatives by the analog of 
Eqs.~(\ref{effdensity})-(\ref{piab-phi}), obtained by 
replacing non-tilded with tilded quantities. In terms of the ``original'' 
$\phi$-fluid, the effective energy-momentum tensor of the ``new'' 
$\tilde{\phi}$-fluid is
\begin{eqnarray}
\tilde{T}_{ab}^{(\tilde{\phi}) } &=& T_{ab}^{(\phi) }
+ \frac{\alpha}{4\pi \phi} \left[ \frac{(1+\alpha)}{\phi} \, \nabla_a \phi 
\nabla_b \phi  \right. \nonumber\\
&&\nonumber\\
&\, & \left. +\frac{(\alpha-2)}{2\phi} \, 
\nabla^e\phi \nabla_e \phi \, 
g_{ab} -\left( \nabla_a \nabla_b \phi -g_{ab} \Box \phi \right) \right] 
\,.\nonumber\\
&&
\end{eqnarray}
We then proceed to find the new fluid  quantities in the comoving frame 
expressed in terms of the original ones. The effective energy density is
\begin{eqnarray} 
\tilde{\rho}^{( \tilde{\phi})} &=& \phi^{-2\alpha} \left[ \rho^{(\phi)} 
- \frac{3 \alpha^2}{8\pi \phi^2} \nabla^e \phi \nabla_e \phi 
\right. \nonumber\\
&&\nonumber\\
&\, & \left. -\frac{\alpha}{4\pi \phi} \left( \Box\phi 
-\frac{ \nabla^a \phi \nabla^b \phi \nabla_a \nabla_b \phi }{ 
\nabla^e \phi \nabla_e \phi }  \right) 
\right] \,.
\end{eqnarray}
Using Eq.~(\ref{effdensity}), one can rewrite it as 
\be
\tilde{\rho}^{( \tilde{\phi})} = \phi^{-2\alpha} \left[ (1-2\alpha) 
\rho^{(\phi)}  - \frac{\alpha(3\alpha + \omega)}{8 \pi \phi^2} 
\, \nabla^e \phi \nabla_e \phi  + \frac{\alpha V}{8 \pi \phi}   
\right] \,.
\ee
The effective heat flux density is
\begin{eqnarray} 
\tilde{q}_a^{( \tilde{\phi})} &=& \phi^{-\alpha} \Bigg\{ q_a^{(\phi)} + 
\frac{ \alpha}{4\pi \phi \sqrt{ -\nabla^e \phi \nabla_e \phi} } 
\Big[ \nabla_a\nabla_c  \phi \nabla^c \phi \nonumber\\
&&\nonumber\\
&\, & - \frac{
\left( \nabla_c \nabla_d \phi \nabla^c  \phi \nabla^d \phi \right)  
 }{   \nabla^e \phi \nabla_e  \phi} \, \nabla_a \phi  \Big] \Bigg\}
\end{eqnarray}
and, using Eq.~(\ref{eq:q}), this turns into 
\be
\tilde{q}_a^{( \tilde{\phi})} = \left( 1-2\alpha \right) \phi^{-\alpha}\,   
q_a^{(\phi)}  \,.
\ee
The effective pressure is 
\begin{eqnarray} 
\tilde{P}^{( \tilde{\phi})} &=& \phi^{-2 \alpha} \left\{ P^{(\phi)} + 
\frac{\alpha}{4\pi \phi} \left[ \frac{(\alpha-2)}{2\phi} \, \nabla^e 
\phi \nabla_e \phi  + \frac{2}{3}\,  \Box \phi \right.\right.\nonumber\\
&&\nonumber\\
&\, & \left.\left. 
+\frac{  \nabla_a \nabla_b \phi  \nabla^a \phi \nabla^b \phi}{  
3 \nabla^e \nabla_e \phi}  \right] \right\} 
\end{eqnarray}
which, using Eq.~(\ref{effpressure}), becomes 
\begin{eqnarray}
\tilde{P}^{( \tilde{\phi})} &=& \phi^{-2 \alpha} \left[ \left( 1-2\alpha 
\right) P^{(\phi)} + 
\frac{\alpha \left(\alpha - \omega -2 \right)}{8 \pi \phi^2}  \, 
\nabla^e \phi \nabla_e \phi \right.\nonumber\\
&&\nonumber\\
&\, & \left.  - \frac{\alpha V}{8 \pi 
\phi}  \right] \,.
\end{eqnarray}
The effective spatial stress tensor is computed as 
\begin{eqnarray} 
\tilde{\Pi}_{ab}^{( \tilde{\phi})} &=& \Pi_{ab}^{(\phi)} 
+ \frac{\alpha}{4\pi \phi} \left[ \frac{ \left( \alpha-2 \right)}{2 \phi} 
\, \nabla^e \phi \nabla_e \phi  + \Box \phi \right] h_{ab}  
\nonumber\\
&&\nonumber\\
&\, &   - \frac{ \alpha}{4\pi \phi} \Big[ \nabla_a \nabla_b \phi \nonumber\\
&&\nonumber\\
&\, & - \frac{ \nabla^c\phi \left( \nabla_a \nabla_c \phi \nabla_b 
\phi + 
\nabla_b \nabla_c \phi \nabla_a \phi\right) }{\nabla^e\phi \nabla_e\phi} 
\nonumber\\
&&\nonumber\\
&\, & + \frac{ \left( \nabla_c  \nabla_d \phi  \nabla^c \phi 
\nabla^d \phi \right) }{  \left( \nabla^e \phi \nabla_e  \phi \right)^2} 
\,  \nabla_a \phi \nabla_b \phi   \Big]  \\
&&\nonumber\\
& = &  
\Pi_{ab}^{(\phi)} + \frac{ \alpha}{4\pi \phi} \left[  \frac{\left( 
\alpha-2\right)}{2 \phi} \, \nabla^e \phi \nabla_e \phi + \Box \phi 
\right] h_{ab} \nonumber\\
&&\nonumber\\
&\, & - \frac{\alpha}{4\pi\phi} \, \nabla_c \nabla_d 
\phi \, {h_a}^c {h_b}^d  \,. \label{annvedi}
\end{eqnarray}
The use of Eq.~(\ref{eq:effPi2}) in Eq.~(\ref{annvedi}) then gives
\begin{eqnarray}
\tilde{\Pi}_{ab}^{( \tilde{\phi})} &=& \left( 1-2\alpha 
\right)\Pi_{ab}^{(\phi)}  \nonumber\\
&&\nonumber\\
&\, & + \frac{\alpha}{8 \pi \phi} \left[ \frac{\left( \alpha - 
\omega-2\right)}{\phi} \, \nabla^e \phi \nabla_e \phi  - V \right] h_{ab} 
\,.\nonumber\\
&&
\end{eqnarray}
The effective anisotropic stresses are simply 
\be
\tilde{\pi}_{ab}^{( \tilde{\phi})} = \left( 1-2\alpha \right) 
\pi_{ab}^{(\phi)}  
\ee
in terms of those associated with the $\phi$-fluid. 

A symmetry transformation~(\ref{newg}), (\ref{newphi}) with $\alpha>1/2$ 
reverses the sign of the heat flux density and of the anisotropic 
stresses. For example, in spherical symmetry a (radial) ingoing energy 
flow will be changed into an outgoing flow by such a transformation. This 
fact is of some interest in the context of inhomogeneous universes 
describing black holes embedded in cosmological backgrounds 
\cite{mylastbook}.

\section{Effective fluid description of $f(R)$ gravity}
\label{sec:6}
\setcounter{equation}{0}

$f(R)$ gravity, which is an extremely popular class of theories used to 
explain the present acceleration of the universe without dark energy 
\cite{CCT}, is described by the action 
\be
S_{f(R)}= \frac{1}{16\pi} \int d^4x \sqrt{-g} \, f(R)   +S^{(m)} 
\,,\label{f(R)action}
\ee
where $f(R)$ is a nonlinear function of the Ricci scalar  
and, as usual,  
$S^{(m)}$ is the action of ordinary matter. It is well known that 
the  gravitational action $ S_{f(R)}$ is equivalent to that of a 
Brans-Dicke  theory  with Brans-Dicke field $\phi =f'(R)$, Brans-Dicke 
coupling $\omega=0$, and 
scalar field  potential \cite{reviews} 
\be
V(\phi) =Rf'(R) -f(R) \bigg|_{R=R(\phi)} \,, \label{f0}
\ee
where $R$ is now a function of the scalar field $\phi=f'(R)$ and a prime 
denotes differentiation with respect to the Ricci scalar $R$. In general, 
the relation $R=R(\phi)$ cannot be inverted explicitly to obtain an 
explicit function $V(\phi)$.

The fourth order vacuum field equations are 
\begin{equation}
f'(R)R_{ab}-\frac{ f(R)}{2} \, g_{ab}=  \nabla_a \nabla_b f'(R) - g_{ab} 
\square f'(R) 
\end{equation}
and can be written as the effective Einstein equations \cite{reviews}
\be
R_{ab}-\frac{1}{2} \, g_{ab} R = 8\pi  \, T_{ab}^{(eff)} 
\,,\label{effectiveEFE}
\ee
where
\begin{eqnarray}
T_{ab}^{(eff)} &=& \frac{1}{8\pi f'(R)} \Big[ \nabla_a \nabla_b f'(R) 
-g_{ab} \Box f'(R)  \nonumber\\
&&\nonumber\\
&\, &  +\frac{ f(R)-Rf'(R)}{2} \, g_{ab} \Big] \,.
\end{eqnarray}
We need the transformation properties
\begin{eqnarray}
&& \nabla^a f' =  f'' \nabla^a R \,, \label{eq:100-1}\\
&&\nonumber\\
&& \nabla_a \nabla_b f'  =  
 f''  \nabla_a \nabla_b R + f'''\nabla_a R \nabla_b R \,, 
\label{eq:100-2}\\
&&\nonumber\\
&& \square f' = f''  \square R + f'''  \nabla^e R \nabla_e R \,. 
\label{eq:100-3}
\end{eqnarray}
We have  $\phi=f'(R)>0$ in order for the graviton to carry positive 
kinetic energy, while we require $f''(R)>0$ to avoid the 
notorious Dolgov-Kawasaki instability 
\cite{DolgovKawasaki, mattmodgrav}. As  a result, the condition that 
$\nabla_a \phi$ be 
timelike implies that $\nabla_a R$ is also timelike and the 
definition~(\ref{4-velocity}) of the effective fluid four-velocity yields
\be
u_a= \frac{ \nabla_a R}{\sqrt{ -\nabla^eR \nabla_eR}} \,.\label{eq:100-4}
\ee
The effective fluid quantities are computed using 
Eqs.~(\ref{eq:100-1})-(\ref{eq:100-4}) in the expressions of the effective 
scalar-tensor fluid, obtaining
\begin{widetext}
\begin{eqnarray}
\rho^{(eff)} &=& \frac{1}{8\pi f'} \Big\{ f'' \left[ \square R 
-\frac{\nabla_a \nabla_b R \nabla^a R \nabla^b R}{ \nabla^e R \nabla_e R} 
 \right] +  \frac{Rf'-f}{2} \Big\} \,,\\
&&\nonumber\\
q_a^{(eff)} &=& \frac{ f''}{8\pi f' \sqrt{-\nabla^e R \nabla_e R} } 
 \left[  \frac{ \left( \nabla_c  \nabla_d R \nabla^c R  \nabla^d R \right)}{   
\nabla^e R \nabla_e R } \, \nabla_a R - \nabla_a \nabla_c R 
\nabla^c R \right] \,,\\
&&\nonumber\\
\Pi_{ab}^{(eff)} &=& h_a^c h_b^d \, \nabla_c \nabla_d f'(R) - \left( 
\square f'(R) + \frac{Rf'-f}{2}  \right)h_{ab} \\
&&\nonumber\\
&=& \frac{1}{8\pi f'} \Big\{  f'' \left[ \nabla_a \nabla_b R +
\frac{ \left( \nabla_a R \nabla_b \nabla_c R +
\nabla_b R \nabla_a \nabla_c R \right) \nabla^c R }{ -\nabla^e R \nabla_e 
R} + \frac{ \left( \nabla_c \nabla_d R    \nabla^c R \nabla^d 
R\right)\nabla_a R \nabla_b R }{ \left(\nabla^e R \nabla_e R\right)^2} 
\right] \nonumber\\
&&\nonumber\\
&\, &  - \left( f'' \square R + f'''  \nabla^e R \nabla_e R  
+ \frac{ Rf'-f }{2} \right) h_{ab} \Big\}\,,\\
&&\nonumber\\
P^{(eff)} &=&  \frac{1}{8\pi f'} \Big[  -\frac{f''}{3} \left(
2 \square R +  \frac{ \nabla_a\nabla_b R \nabla^a R \nabla^b R }{ \nabla^e 
R \nabla_e R}\right) -\left( f'''    \nabla^e R \nabla_e R +  \frac{ 
Rf' - f}{2} \right) \Big] \,,\\
&&\nonumber\\
\pi_{ab}^{(eff)} &=&  \frac{f''}{8\pi f'} \Big[ \nabla_a  \nabla_b R
- \frac{ \left( \nabla_a R \nabla_b \nabla_c R + \nabla_b R 
 \nabla_a \nabla_c R \right)  \nabla^c R}{  \nabla^e R \nabla_e R} 
+  \frac{ \left( \nabla_c \nabla_d R  \nabla^c R \nabla^d R \right)}{
\left( \nabla^e R \nabla_e R \right)^2} \,   \nabla_a R  \nabla_b R 
\nonumber\\
&&\nonumber\\
&\, & +\frac{1}{3} \left( \frac{
\nabla_c\nabla_d R \nabla^c R \nabla^d R }{\nabla^eR \nabla_e R} 
- \Box R \right) \, h_{ab} \Big]  \,. 
\end{eqnarray}
\end{widetext}
Finally, the trace of the effective stress-energy tensor is
\begin{eqnarray}
&& T^{(eff)} = -\rho^{(eff)} + 3P^{(eff)} \nonumber\\
&&\nonumber\\
&  & = \frac{1}{8\pi f'}  
\Big[  -3 \left( f'' \Box R + f''' \nabla^e R \nabla_e R\right) + 
2\left(f- Rf' \right) \Big] \,.\nonumber\\
&&
\end{eqnarray}
It is thus demonstrated that, in general, the terms generated by a 
nonlinear function $f(R)$ in the gravitational Lagrangian are equivalent 
to an imperfect fluid when writing the field equations as the effective 
Einstein equations~(\ref{effectiveEFE}). In special geometries, this 
imperfect fluid reduces to a perfect fluid 
dubbed ``curvature fluid'' \cite{CapozzielloLaurentisLambiase, 
NojiriOdintsov, Zimdahl}. This is 
the case for the FLRW geometry \cite{CapozzielloLaurentisLambiase, 
NojiriOdintsov, Zimdahl},  
for the Lorentzian version of a Hawking  wormhole \cite{Culetu}, and for a 
Witten bubble spacetime solution \cite{Culetu}.

\section{Discussion and conclusions}
\label{sec:7}
\setcounter{equation}{0}

The fluid equivalent of the Brans-Dicke-like scalar field $\phi$ of 
scalar-tensor gravity in the Jordan frame has been worked out in detail, 
completing and extending the work of Ref.~\cite{Pimentel89}. 
The field equations~(\ref{BDfe1}) and (\ref{BDfe2}) can be regarded as 
effective Einstein equations and the terms originating from $\phi$ and its 
derivatives, relegated to the right hand side, can always be interpreted 
as an effective fluid. Contrary to the case of a canonical scalar field 
minimally coupled to the curvature, which is definitely a matter field of 
non-gravitational nature and is equivalent to a perfect fluid, the 
effective fluid corresponding to a Brans-Dicke-like field is an imperfect 
one (except for special circumstances in highly symmetric 
geometries---but we refer to the general situation here). As expected, 
since the effective fluid is generated by a purely scalar degree of 
freedom, it is irrotational. Dissipation in fluids is important and it is 
related to the stability of star models \cite{asteroseismology}, while 
dissipative fluids are also the subject of a vast literature related to 
the AdS/CFT correspondence \cite{AdSCFT}. The discussion and formulas 
presented here should find applications when modified gravity is discussed 
in conjunction with these areas of research.

Contrary to what we have done here, in principle one could have started 
out with the 
Einstein frame representation of scalar-tensor gravity, obtained by the 
conformal transformation and nonlinear field 
redefinition $\left( g_{ab}, \phi \right) \rightarrow \left( 
\tilde{g}_{ab}, \tilde{\phi} \right)$ with
\begin{eqnarray}
\tilde{g}_{ab} & = & \phi \, g_{ab} \,,\\
&&\nonumber\\
d\tilde{ \phi} & = &  
\sqrt{ \frac{|2\omega(\phi)+3|}{16\pi G} } \, \frac{d\phi}{\phi} \,,
\end{eqnarray}
which is completely different from the symmetry~(\ref{newg}) and 
(\ref{newphi}) discussed in Sec.~\ref{sec:5}. The Einstein frame version 
of the scalar-tensor action~(\ref{STaction}) is 
\begin{eqnarray}
S_{ST} &=& \int d^4 x \, \sqrt{-\tilde{g}} \left[ \frac{ \tilde{R}}{16\pi 
G} -\frac{1}{2} \, \tilde{g}^{ab} \tilde{\nabla}_a \tilde{\phi}
\tilde{\nabla}_b \tilde{\phi} -U( \tilde{\phi}) \right. \nonumber\\
&&\nonumber\\
&\, & \left. + \frac{   {\cal L}^{(m)} \left[ \phi^{-1} 
\tilde{g}_{cd}, \psi^{(m)} \right] }{ \phi^2(\tilde{\phi}) }  \right] \,,
\end{eqnarray}
where 
\be
U(\tilde{\phi}) = \frac{ V(\phi)}{\phi^2} \left|_{ \phi=\phi( 
\tilde{\phi}) } \right. 
\ee
and $\psi^{(m)}$ collectively denotes the matter fields.
The Lagrangian density of the new scalar $\tilde{\phi}$ has canonical form 
save for the fact 
that it couples explicitly to all other forms of matter except 
conformally invariant matter \cite{FujiiMaeda, mySTbook}. If one considers 
(electro-)vacuum scalar-tensor gravity, the Einstein frame scalar is 
equivalent to a perfect fluid. In this case (but not in the general case 
of non-conformal matter),  the transformation from Jordan 
to Einstein frame changes an equivalent imperfect fluid into a perfect 
one, and {\em vice-versa.} (The behaviour of more general perfect and 
imperfect fluids under conformal transformations is discussed in 
Ref.~\cite{Clarkson}.)

Finally, let us discuss the implications of the work presented here 
for the different, long-standing problem of finding a Lagrangian 
description of a dissipative imperfect fluid. It is notoriously difficult 
to give a Lagrangian or Hamiltonian description of dissipative systems 
\cite{diss-fluids}, except for 
simplistic models of friction in point particle mechanics 
\cite{Goldstein}. The problem is even more difficult in fluid 
mechanics \cite{diss-fluids}. By reversing our original problem of finding 
an effective fluid description of a scalar field theory, we are able to 
provide  a very limited answer: an irrotational imperfect fluid with  
energy-momentum tensor~(\ref{imperfectTab}) can be given a scalar field 
description and, therefore, a full Lagrangian description, if its 
irrotational four-velocity field can be written in the 
form~(\ref{4-velocity}) for a 
suitable scalar $\phi$. Needless to say, this is an extremely restrictive 
condition which makes the answer useless for most practical purposes 
because, in general, one cannot integrate Eq.~(\ref{4-velocity}) to 
determine the velocity potential $\phi$, but this condition has been 
usefully implemented using a Lagrange multiplier (a second scalar field) 
in more complicated theories \cite{LimSawickiVikman,MirzagholiVikman}.  A 
much simpler 
problem occurs 
by restricting oneself to specific spacetimes with a high degree of 
symmetry. If the scalar 
field $\phi$ is forced to depend on just one of the four coordinates, 
which can happen only in highly symmetric spaces because $\phi$ is a 
matter source, then Eq.~(\ref{4-velocity}) simplifies considerably. This 
is the 
case, for example, of spatially homogeneous and isotropic 
FLRW cosmology. In a FLRW space 
with line element
\be
ds^2 =-dt^2 +a^2(t) \left[ \frac{dr^2}{1-kr^2} +r^2 \left( d\theta^2 
+\sin^2 
\theta d\varphi^2 \right) \right] \,,
\ee
the gravitating scalar field depends only on time, $\phi=\phi(t)$, and 
then the fluid four-velocity simplifies to $u_{\mu} =-\delta_{0\mu}$. Even 
in this case, however, integrating Eq.~(\ref{thetaScalar}) with 
$\theta=3H\equiv 3\dot{a}/a$  to determine  $\phi(t)$ is not a trivial 
task because this equation is nonlinear for a general potential 
$ V(\phi)$. The conclusion is that the scalar field-fluid correspondence 
does not allow for significant progress in the problem of the Lagrangian 
description of dissipative fluids. An alternative approach to 
dissipation, obtained by resorting 
to a modification of gravity different from the scalar-tensor 
prescription and doing away with dark energy, is discussed in 
Ref.~\cite{Lazoetal2017}. Apart from  
the approach to a Lagrangian description of dissipation, the 
correspondence  between 
Brans-Dicke-like scalar field and fluid is 
now clarified in the important situations where a scalar field   
appears  in the modelling of cosmology and stellar interiors in the 
context of modified (scalar-tensor and $f(R)$) gravity.

As a last remark, we mention that we presented a general 
formalism without committing to any specific geometry. Two applications 
to specific spacetime geometries would be particularly interesting: the 
first is the case of perturbed 
Friedmann-Lema\^itre-Robertson-Walker universes, and the second is the 
perturbation of black hole spacetimes. In the first case, perturbations 
of universes 
filled with an imperfect fluid have been studied in the literature (see 
\cite{MalikWands} for a review). This description may be adapted 
to scalar-tensor gravity, with the addition of a matter fluid to the 
picture. 
This application of the imperfect fluid formalism presented here 
necessarily involves many details and will be presented elsewhere. In the 
second case, perturbations of black holes in scalar-tensor gravity have 
been studied, but perturbations in the presence of an imperfect 
fluid in general relativity are less clear and will also require a 
separate analysis.  Likewise, the imperfect fluid corresponding to 
modified gravity constitutes a form of non-adiabatic dark energy quite 
different from the standard dark energy models, which 
would give rise to non-adiabatic perturbations. Non-adiabatic dark energy 
has been considered in various works ({\em e.g.}, 
Ref.~\cite{non-adiabaticDE}), 
but the detailed relations with the present imperfect fluid formalism are  
still missing and will be explored elsewhere.

\begin{acknowledgments}

We thank Thomas Gobeil and Shawn Belknap-Keet for discussions and a 
referee for useful insight and references. This work is supported, in 
part, by the Natural Sciences and Engineering Research Council of Canada 
(Grant No.~2016-03803 to V.F.).

\end{acknowledgments}

%\bibliography{simplified}

\end{document}